\documentclass[twocolumn,superscriptaddress,prl]{revtex4-1}

\usepackage{color}

\usepackage{amsmath}
\usepackage{amssymb}
\usepackage{graphicx}

\begin{document}

\title{Electronic magnetization of a quantum point contact \\
measured by nuclear magnetic resonance}

\author{Minoru Kawamura}
\email{minoru@riken.jp}
\affiliation{RIKEN Center for Emergent Matter Science, Wako 351-0198, Japan}

\author{Keiji Ono}
\affiliation{RIKEN Center for Emergent Matter Science, Wako 351-0198, Japan}

\author{Peter Stano}
\affiliation{RIKEN Center for Emergent Matter Science, Wako 351-0198, Japan}

\author{Kimitoshi Kono}
\affiliation{RIKEN Center for Emergent Matter Science, Wako 351-0198, Japan}

\author{Tomosuke Aono}
\affiliation{Department of Electrical and Electronic Engineering, Ibaraki University,  Hitachi 316-8511, Japan}

\date{\today}

\begin{abstract}
We report an electronic magnetization measurement of a quantum point contact (QPC)
based on nuclear magnetic resonance (NMR) spectroscopy.
We find that NMR signals can be detected by measuring the QPC conductance
under in-plane magnetic fields.
This makes it possible to measure, from Knight shifts of the NMR spectra,
the electronic magnetization of a QPC containing only a few electron spins.
The magnetization changes smoothly with the QPC potential barrier height
and peaks at the conductance plateau of 0.5 $\times$ $2e^2/h$.
The observed features are well captured by a model calculation assuming a smooth potential barrier,
supporting a no bound state origin of the 0.7 structure.
\end{abstract}

\maketitle

Quantum point contact (QPC) is a short one-dimensional (1D) channel
connecting two electron reservoirs.
Its conductance is quantized to integer multiples of $2e^2/h$,
where $e$ is electron charge and $h$ is Planck's constant\cite{vanWees, Wharam}.
The conductance quantization is well understood within a model of non-interacting electrons\cite{Buttiker}.
However, experiments have shown an additional conductance feature,
a shoulder-like structure at around 0.7 $\times$ $2e^2/h$
termed as  0.7 structure\cite{Thomas1996, Thomas1998}.
Despite the simplicity of a QPC, a comprehensive understanding
of the 0.7 structure is still lacking \cite{ Wang, Reilly, Matveev, Cronenwett, Meir, Rejec,
Sloggett, Bauer, Yoon, DiCarlo, Iqbal, Brun, Smith, Micolich}.

Theories proposed to explain the 0.7 structure
can be discriminated according to their predictions on the electron spin arrangement,
which include spontaneous spin polarization\cite{Wang, Reilly},
antiferromagnetic Wigner crystal\cite{Matveev},
Kondo screening\cite{Cronenwett, Meir, Rejec},
and local spin fluctuations accompanied by van Hove singularity\cite{Sloggett, Bauer}.
Especially in the Kondo scenario, the existence of a localized magnetic moment in the QPC
is an inevitable ingredient.
On one hand, early experiments observing Fano resonances suggested
such presence of a local single spin trapped in a bound state regardless of magnetic fields\cite{Yoon}.
On the other hand, an experiment measuring compressibility
contradicts such bound state formation\cite{Smith}.
Thus, the degree of spin polarization of a QPC is one of the central issues 
to understand the origin of the 0.7 structure.

However, most experiments\cite{Thomas1996, Thomas1998, Yoon, DiCarlo, Iqbal, Brun} to date
have focused on transmission properties, without the QPC spin polarization being addressed directly.
Despite the recent progress in magnetic sensors\cite{Grinolds},
the magnetization measurement of a QPC containing only a few electrons is still very challenging.
Recently, small magnetizations of  two-dimensional electron systems (2DESs) embedded in GaAs
have been measured\cite{Kumada, Tiemann, Stern}
by combining techniques of current-induced nuclear spin polarization
\cite{Kane, Wald, Dixon, Kronmuller, MKapl}
and resistance (conductance) detection of nuclear magnetic resonance (NMR) signals
of Ga and As nuclei\cite{Kronmuller, MKapl, Desrat}.
Because of the hyperfine interaction between electronic and nuclear spins,
an electronic magnetization produces an effective magnetic field for nuclei,
resulting in the shift of the NMR frequency, the Knight shift.
From the Knight shift, the electronic magnetization can be determined\cite{Slichter}.

A recent transport experiment by Ren {\it et al.}\cite{Ren} suggests
such influence of nuclear spins on the QPC conductance.
They observed  hysteresis in the source-drain voltage dependence of the differential conductance
under magnetic fields, and attributed its origin to the dynamical nuclear spin polarization (DNSP) 
induced in the QPC.
However, NMR or other direct evidence showing involvement of nuclear spins has not been presented so far.
NMR signal detection in the QPC conductance would constitute 
a novel experimental technique to probe spin properties of QPCs or nanowires\cite{Cooper, Stano}.

In this Letter, we report an electronic magnetization measurement of a QPC
defined in a GaAs/AlGaAs heterostructure based on NMR spectroscopy.
We find that the QPC differential conductance changes
when the frequency of an applied oscillating magnetic field matches
the NMR  frequencies of $^{69}$Ga, $^{71}$Ga, and $^{75}$As.
The resistive detection of the NMR signals allows us
to measure the electronic magnetization of the QPC from the Knight shifts of the NMR spectra.
The Knight shift measurements are conducted at the QPC conductance
between 0 and  $2e^2/h$ by tuning gate and source-drain voltages.
The magnetization changes smoothly with the QPC potential barrier height
and peaks at the conductance plateau of 0.5 $\times$ $2e^2/h$.
The observed features are well captured by a model calculation
assuming a smooth potential barrier without a bound state formed.
Apart from the demonstration of a new technique to measure a magnetization of only a few electrons,
the absence of a bound state in the QPC is our main conclusion,
directly relevant for the understanding of the 0.7 structure.

\begin{figure}[t]
	\begin{center}
		\includegraphics[width=78.2mm]{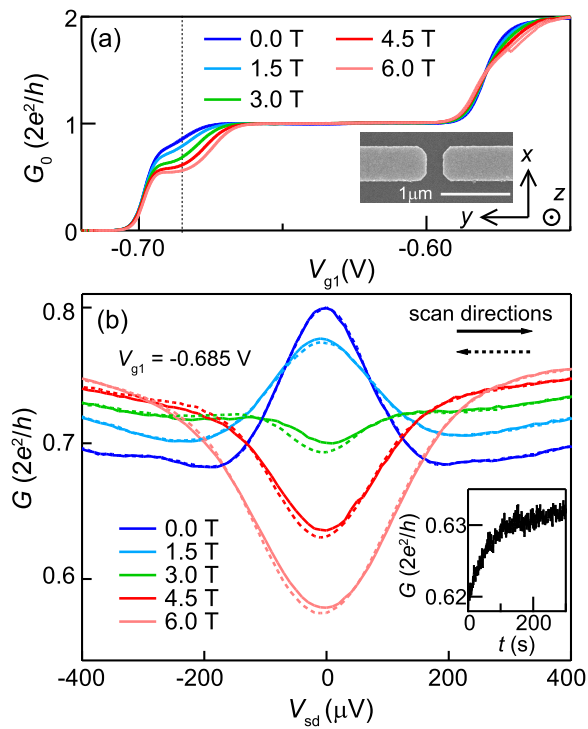}
	\end{center}
		\caption{
		\label{hysteresis}
		(color online)
		(a) Linear conductance $G_0$ as a function of $V_{\rm g1}$ ($V_{\rm g2}$ = $-$1.4 V) 
		at $B$ = 0, 1.5, 3, 4.5, and 6 T, applied along the $x$ direction.
		Inset shows a scanning electron microscope image of the device.
		(b) Differential conductance $G$ as a function of source-drain bias voltage $V_{\rm sd}$
		at $V_{\rm g1}$ = $-$0.685 V [indicated by a dashed line in (a)] 
		under the same magnetic fields as in (a).
		The solid (dashed) curves are measured by scanning $V_{\rm sd}$ 
		in the positive (negative) direction at a rate of 5.6 $\mu$V/s.
			Inset shows the time dependence of $G$
			at $B$ = 4.5 T and $V_{\rm g1}$ = $-0.685$ V
			after an instantaneous change of $V_{\rm sd}$
			from  0 to $-50$ $\mu$V.
			A slightly different value compared to the one in (b) for the same parameters,
			$B$ = 4.5 T and $V_{\rm sd}$ = $-$50 $\mu$V,
			arises due to a remaining DNSP
			created at large $|V_{\rm sd}|$ during the $V_{\rm sd}$ scan in (b).
		}
\end{figure}

QPCs  studied in this work are fabricated from a wafer of GaAs/Al$_{0.3}$Ga$_{0.7}$As
single heterostructure with a 2DES at the interface.
The mobility and sheet carrier density of the 2DES at 4.2 K
are 110 m$^{2}$/Vs and 2.2 $\times$ 10$^{15}$ m$^{-2}$, respectively.
A QPC is defined electro-statically applying negative voltages ($V_{\rm g1}$, $V_{\rm g2}$)
to a pair of Au/Ti gate electrodes patterned on the surface of the wafer.
All data presented here are measured on a QPC with lithographic dimensions of
 300 nm length and 250 nm width [inset of Fig.~\ref{hysteresis}(a)],
in a dilution refrigerator at the mixing chamber temperature of 20 mK.
The external magnetic field $B$ is applied parallel to the 2DES plane
along the current flowing direction [$x$ direction in the inset of Fig. 1(a)]
to avoid orbital effects and quantum Hall edge channels.
The differential conductance $G = dI/dV_{\rm sd}$
(where $I$ is the current and $V_{\rm sd}$ is the source-drain bias voltage)
 is measured using a standard lock-in technique with a typical excitation voltage  of 20 $\mu$V at 118 Hz.
A single-turn coil is wound around the device to apply radio-frequency
oscillating magnetic field $B_{\rm rf}$.

The QPC shows a typical conductance quantization behavior.
Figure \ref{hysteresis}(a) shows linear conductance $G_0 = G(V_{\rm sd} = 0)$
as a function of gate voltage $V_{\rm g1}$.
In addition to quantized conductance plateaus, the 0.7 structure is observed at zero magnetic field,
developing into a plateau of  0.5 $\times$ $2e^2/h$ at high magnetic fields.
A zero-bias conductance peak accompanying the 0.7 structure is observed clearly
in the $G$-$V_{\rm sd}$ curve at $B$ = 0 T [Fig.~\ref{hysteresis}(b)].
With increasing $B$, the zero-bias conductance peak is suppressed and turns into a dip above $B$ = 3 T.

Hysteresis is observed in the $G$-$V_{\rm sd}$ curves when $V_{\rm sd}$ is scanned slowly
 (5.6 $\mu$V/s) in the positive and negative directions [Fig.~\ref{hysteresis}(b)].
The hysteresis is seen only at finite magnetic fields.
Typical time scale to develop the hysteresis is measured at $B$ = 4.5 T by recording $G$
after an instantaneous change of $V_{\rm sd}$ from 0  to $-50$ $\mu$V 
[Inset of Fig.~\ref{hysteresis}(b)].
The value of $G$ continues to change over a period of 200 s.
This time scale is consistent with nuclear spin relaxation or polarization times
reported in GaAs-based devices\cite{Kronmuller, Ren, MKapl, MKfrac,  kawamura2013}.
Similarly as concluded in the earlier work\cite{Ren}, we interpret the slow change in $G$
as the first indication for the DNSP in the QPC.

\begin{figure}[t]
	\begin{center}
		\includegraphics[width=78.0mm]{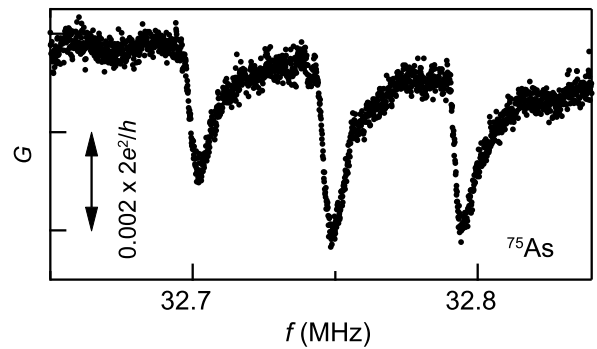}
	\end{center}
		\caption{
		\label{cwNMR}
		Differential conductance $G$ as a function of frequency $f$ of $B_{\rm rf}$
		at $B$ = 4.5 T, $V_{\rm g1}$ = $-0.685$ V, and $V_{\rm sd}$ = $-$50 $\mu$V.
		$B_{\rm rf}$ is applied  perpendicular to $B$ [$y$ direction in the inset of Fig. 1(a)].
		$f$ is scanned at a rate of  0.128 kHz/s. 
		Data of 10 subsequent measurements are averaged to improve the signal to noise ratio.
		}
\end{figure}

To confirm the nuclear spin origin of the observed slow change in $G$,
we perform the NMR spectroscopy experiment.
Scanning the frequency $f$ of $B_{\rm rf}$, we observe decreases in $G$ when $f$ matches 
the NMR frequency of $^{75}$As (gyromagnetic ratio $\gamma$ = 45.82 rad MHz/T) [Fig.~\ref{cwNMR}].
The obtained $G$-$f$ curve represents the NMR spectrum of $^{75}$As,
split into three dips due to the electric quadrupole interaction\cite{suppl}.
We observe signals at resonances of  $^{69}$Ga and  $^{71}$Ga,
as well as  analogous behavior in four other QPC devices (not shown).
These observations clearly show that the DNSP is induced in the QPC
and that its changes are measured by monitoring the QPC conductance.

\begin{figure*}[t]
	\begin{center}
		\includegraphics[width=155mm]{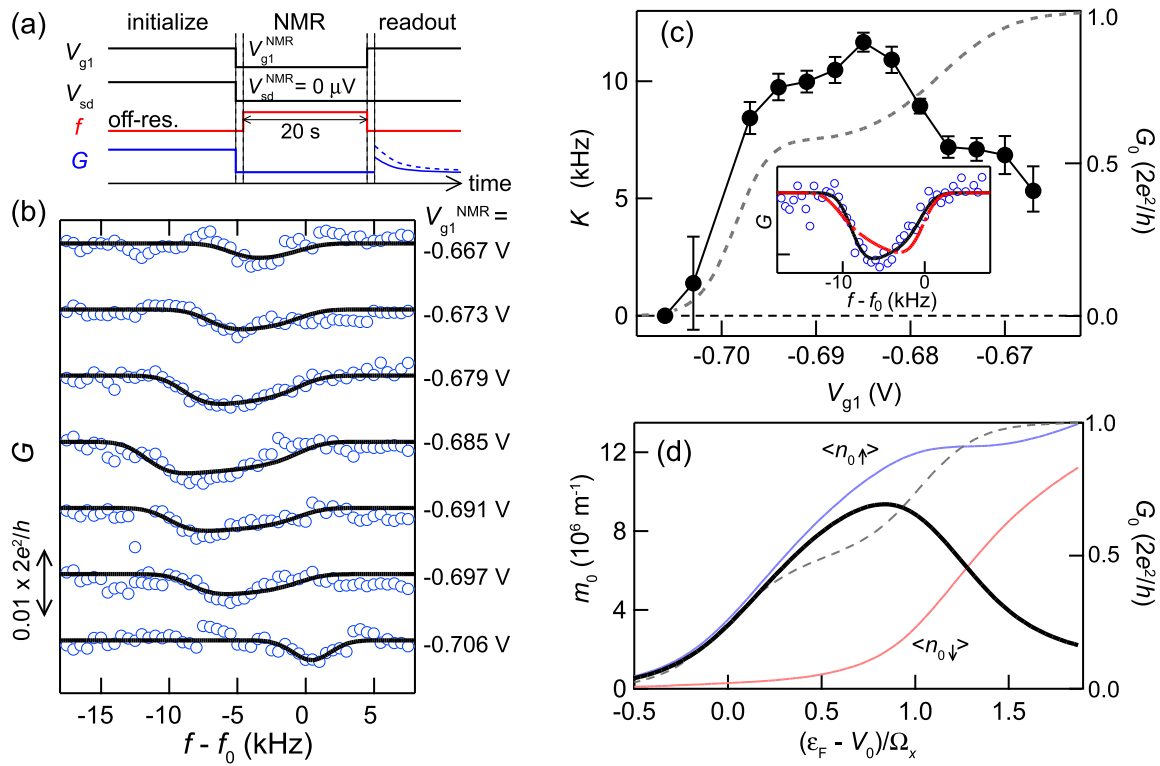}
	\end{center}
		\caption{
		\label{Knightshift}
		(color online)
		(a) Schematic sequence for the pump-probe experiment.
		(b) NMR spectra of $^{75}$As for $V_{\rm sd}^{\rm NMR}$ = 0 $\mu$V
		and various gate voltages $V_{\rm g1}^{\rm NMR}$, as indicated.
		NMR signals corresponding to the transition between 
		nuclear spin states $|I_z = \pm 1/2 \rangle$ are shown.
		Solid curves are the fitting results. Data are offset vertically for clarity.
		(c) Knight shift $K$ plotted as a function of gate voltage $V_{\rm g1}$.
		Linear conductance $G_0$ is plotted by a dotted curve referring to the right axis.
			Inset shows fitting results of the NMR data for $V_{\rm g1}^{\rm NMR}$ = $-0.679$ V
			assuming 3D (dashed) and 2D (solid) hard-wall confinement potentials.
		(d) Calculated magnetization density at the QPC center $m_0$
		plotted as a function of potential barrier height $V_0$.
		The red and blue curves depict spin densities  $\langle n_{0, \uparrow} \rangle$
		and  $\langle n_{0, \downarrow} \rangle$, respectively.
		Calculated conductance $G_0$ is plotted by a dotted curve referring to the right axis.
		}
\end{figure*}

Having established the method to probe the NMR spectra in transport,
we now use it to determine the electronic magnetization of the QPC.
To this end, we perform the following pump-probe experiment [Fig.~\ref{Knightshift}(a)].
First, nuclear spins are initialized by inducing DNSP under
a relatively large bias voltage $V_{\rm sd}$  = $-$300 $\mu$V at $V_{\rm g1}$ = $-0.685$ V.
Then, $V_{\rm sd}$ is set to 0 $\mu$V and 
the QPC is tuned to a state of interest by setting the gate voltage to $V_{\rm g1}^{\rm NMR}$
for a period of time (22 s), during which the frequency of $B_{\rm rf}$ is set to $f$ for 20 s\cite{Vac}.
Finally, changes in the DNSP are read out by recording $G$ 
with a small ac voltage excitation (20 $\mu$V, 118 Hz)
at $V_{\rm g1}$ = $-0.685$ V and $V_{\rm sd}$ = 0 $\mu$V.
The observed values of $G$ at the beginning of the readout step reflect
how much are the nuclear spins depolarized by $B_{\rm rf}$.
Repeating this procedure with different $f$,
we obtain an NMR spectrum for a gate voltage $V_{\rm g1}^{\rm NMR}$
as shown in Fig.~\ref{Knightshift}(b).

The bottom data of Fig.~\ref{Knightshift}(b) is obtained by depleting
electrons from the QPC during the $B_{\rm rf}$ application.
Therefore, this spectrum is not affected by electrons, 
and has a rather sharp dip at $f_0$ =  32.755 MHz,
the frequency corresponding to the transition between the nuclear spin states $|I_z = \pm1/2 \rangle$.
As $V_{\rm g1}^{\rm NMR}$ is increased, 
the NMR induced dips are shifted toward negative frequencies and broadened.
These shifts are the Knight shifts due to the electronic magnetizations in the QPC.

We now evaluate the magnitude of the Knight shifts
by taking the spatial electron distribution into account.
Extending  earlier works\cite{Kumada, Tiemann, Khandelwal}, 
we adopt a model of electrons confined in the $y$ and $z$ directions
with a transverse wave function  $\psi (y, z)$.
The Knight shift for an As nucleus at  position ($y, z$) can be written as 
$\delta f_{\rm K}(y, z)  = \alpha_{\rm As} m_z |\psi (y, z)|^2$,
where $\alpha_{\rm As}$   = $-2.1\times$ 10$^{-22}$ kHz m$^3$
is the hyperfine coupling coefficient\cite{suppl},
and $m_z \equiv n_{\uparrow} -n_{\downarrow}$
is 1D electronic magnetization density defined as the difference in 1D spin densities.
We make a standard assumption\cite{Khandelwal} that nuclear spins are depolarized
by the rf-magnetic field according to the detuning from the resonance
$\delta f = f - (f_0 + \delta f_{\rm K})$ with a Gaussian profile
$\exp (- \delta f^2/2\gamma^2)$, where $f_0$ and $\gamma$ 
are the NMR frequency and the spectrum width without the influence of the Knight shift, respectively.
Such depolarizations induce the change in the electron Zeeman energy 
which is given by an integral of local nuclear spin depolarization multiplied by electron distribution.
Since these changes are small, we may expand the QPC conductance,
which is a function of the electron Zeeman energy, and get for its change
\begin{equation}
	\delta G(f) = A\int dydz \, \exp (- \delta f^2/2\gamma^2) |\psi (y, z)|^2,
	\label{fittingfunc}
\end{equation}
with $A$ an unknown proportionality coefficient.
To evaluate Eq.~(\ref{fittingfunc}), we
approximate the transverse wave function by the one of a
two-dimensional (2D) hard-wall confinement,
$\psi (y, z) \propto \cos(\pi y/w_y) \cos(\pi z/w_z)$
with confinement widths $w_y$ = (65 $\pm$ 5) nm and $w_z$ = (18 $\pm$ 3) nm\cite{parameters}.
The Knight shift becomes $\delta f_{\rm K} (y, z) = - K\cos^2(\pi y/w_y) \cos^2(\pi z/w_z)$
with a parameter $K$ proportional to $m_z$ via $K =  - \alpha_{\rm As} m_z |\psi (0, 0)|^2$.
The experimental data in Fig.~\ref{Knightshift}(b) are fitted to Eq.~(\ref{fittingfunc})
using $K$ and $A$ as fitting parameters
with $f_0$ = 32.755 MHz and $\gamma$ = 1.36 kHz determined
from the data measured at the depletion  configuration ($V_{\rm g1}^{\rm NMR}$  = $-0.706$ V).
As seen in the figure, the agreement of the data 
and the model fitted for each curve is excellent.

We now consider an alternative fit, assuming that the QPC  transport occurs through 
a three-dimensionally (3D) confined electronic state
$\psi (x, y, z) \propto  \cos(\pi x/w_x) \cos(\pi y/w_y) \cos(\pi z/w_z)$.
A representative result, using an analog of Eq.~(\ref{fittingfunc}), is given
in the inset of Fig.~\ref{Knightshift}(c) and shows a much worse compatibility
with the data. We find that such discrepancy is not sensitive to the confinement details.
As especially well visible for large Knight shifts,
the data show skewed line shape, with steep (gentle) slopes on the low (high) frequency side.
This is systematically reproduced by 2D confinement models, unlike 3D ones
(see the Supplemental Material\cite{suppl}).

In Fig.~\ref{Knightshift}(c), $K$ is plotted as a function of $V_{\rm g1}$.
A finite $K$ emerges near the conductance onset
and increases steeply as the conductance is increased to 0.5 $\times$ $2e^2/h$.
It keeps increasing gradually with increasing $V_{\rm g1}$
even in the conductance plateau region of 0.5 $\times$ $2e^2/h$.
As $V_{\rm g1}$ is increased further, $K$ turns to decrease accompanied by 
a rise of conductance from 0.5 $\times$ $2e^2/h$.
As a result, a peak in $K$ is formed at the high gate-voltage end of the conductance plateau.
Using the relation between $K$ and $m_z$,
the observed maximum value $K$ = (11.7 $\pm$ 0.5) kHz
corresponds to  $m_z$ = (16.5 $\pm$ 4.5) $\times$ 10$^{6}$ m$^{-1}$.

We now show that the observed features are well reproduced by a model calculation.
We model a QPC by a 1D tight-binding Hamiltonian,
\begin{equation}
	H = \sum_{j, \sigma} \epsilon_{j, \sigma} c_{j, \sigma}^{\dagger} c_{j, \sigma} 
		- t \sum_{j, \sigma} c_{j, \sigma}^{\dagger} c_{j+1, \sigma} 
		+ \sum_{j}U_j n_{j, \uparrow} n_{j, \downarrow}.
\end{equation}
Here $c_{j, \sigma}^{\dagger}$ creates an electron with spin $\sigma  \;(\sigma=\uparrow,\downarrow)$
at the $j$-th site of the tight-binding chain which has a hopping amplitude $t$.
We assume a short-range Coulomb interaction represented by the on-site Coulomb energy $U_j$.
The potential energy and the Zeeman energy are included in the on-site energy,
$\epsilon_{j, \uparrow / \downarrow} = \epsilon_j  \pm  g \mu_{\rm B} B/2$,
with the Bohr magneton $\mu_{\rm B}$ and the electron $g$-factor $g$.
We assume a smooth parabolic potential barrier at the QPC center with
a height $V_0$ and a curvature $\Omega_x$.
The interaction term is treated by a mean-field approximation neglecting spin fluctuations.
Then the mean-field spin density $\langle n_{j,\sigma} \rangle$ is determined by
a self-consistent Green's function method\cite{Datta},
where the on-site energy $\epsilon_{j,\sigma}$ is shifted by $U_j \langle n_{j,\bar{\sigma}} \rangle$
with $\bar{\sigma}$, the opposite spin to $\sigma$.
We calculate the magnetization density profile
$m_j =  \langle n_{j,\uparrow}  - n_{j,\downarrow} \rangle $ and the QPC conductance $G_0$.
The values of $U_j$ and $\Omega_x$ are determined 
from the conductance measurement data\cite{suppl}.

The thick solid curve in Fig.~\ref{Knightshift}(d) depicts the calculated magnetization density
at the QPC center $m_0 = m_{j =0}$
 as a function of $V_0$,
resembling the observed $V_{\rm g1}$ dependence of $K$ in Fig.~\ref{Knightshift}(c).
According to the calculation,  the increase in $m_0$ accompanied by
the emergence of the conductance corresponds
to the increase in the number of  up-spin electrons in the QPC.
The value of $m_0$ starts to decrease when down-spin electrons begin to populate the QPC,
lifting $G_0$ from 0.5 $\times$ $2e^2/h$.
The gradual increase in $m_0$ in the 0.5 $\times$ $2e^2/h$ plateau region is also reproduced.
The maximum value of the calculated magnetization density $m_0$  = 9.3 $\times$ 10$^6$ m$^{-1}$
roughly agrees with the value determined from the Knight shift.
Spin polarization $P = \langle n_{0,\uparrow} -  n_{0,\downarrow} \rangle/
\langle n_{0,\uparrow} + n_{0,\downarrow} \rangle$ reaches 70.0 \% 
where $m_0$ is maximal.
Distribution of $m_j$ has a bell-shaped profile and extends over a length of about 100 nm
around the QPC center\cite{suppl}.

The gradual change in $m_0$ reflects the fact that
the local density of states is continuous at the QPC center unlike in a quantum dot.
We therefore attribute  the observed gradual change in $K$
to be  consistent with a QPC model without any bound states.
This contradicts earlier observations claiming
that a single electron spin is trapped in a bound state formed in the QPC\cite{Yoon}.
We estimate\cite{suppl} that
the observed magnitude of the magnetization density corresponds to
the total magnetic moment $(1.65 \pm 0.45)$ in the QPC,
exceeding the single-electron-spin magnetic moment which a bound state can support.
Our measurement results of the NMR line shapes,  the gradual change of $K$,
and the magnetic moment values
are consistent with a QPC model without bound states, such as Refs. \cite{Sloggett, Bauer},
which predicts a smooth increase of the magnetization without saturation
upon increasing the magnetic field.

In summary,
we find that the NMR signals can be detected
by measuring the QPC conductance under in-plane magnetic fields.
The resistive detection  makes it possible to measure
the electronic magnetization of the QPC from the Knight shifts of the NMR spectra.
The electronic magnetization changes smoothly with the gate voltage
and peaks at the conductance plateau of 0.5 $\times$ $2e^2/h$.
The gate voltage dependence of the Knight shift
is well explained  by a model calculation assuming a smooth potential barrier,
supporting a no bound state origin of the 0.7 structure.

This work was supported partially by Grant-in-Aid for Scientific Research (No. 24684021)
from JSPS, Japan.
We thank T. Machida for valuable discussions.


\begin{thebibliography}{99}
\bibitem{Vac} 
	The small ac voltage $V_{\rm ac}$ (20 $\mu$V, 118 Hz) was not turned off
	during this period of 22 s. Although the small $V_{\rm ac}$ might induce the DNSP,
	its amplitude is much smaller than that 
	induced by $V_{\rm sd}$ = $-300 \mu$V at the first step.
\bibitem{parameters}
	The parameter $w_z$ = (18 $\pm$ 3) nm
	is typical for the interface of GaAs/AlGaAs\cite{Ando}.
	The parameter $w_y$ = (65 $\pm$ 5) nm, which is related to the inter-subband  energy,
	is determined from the gate voltage width of the first conductance plateau.
\end{thebibliography}
\end{document}